\documentclass[aps,prb,twocolumn,groupedaddress,showpacs]{revtex4}

\usepackage{amssymb}

\usepackage{graphicx}

\begin{document}

\title{Using Qubits to Measure Fidelity in Mesoscopic Systems}

\author{G.B.\ Lesovik$^{a,b}$, F.\ Hassler$^{b}$, and G.\ Blatter$^{b}$}

\affiliation{$^{a}$L.D.\ Landau Institute for Theoretical Physics RAS,
   117940 Moscow, Russia}

\affiliation{$^{b}$Theoretische Physik, ETH-H\"onggerberg, CH-8093
   Z\"urich, Switzerland}

\date{\today}

\begin{abstract}
  We point out the similarities in the definition of the `fidelity' of a
  quantum system and the generating function determining the full counting
  statistics of charge transport through a quantum wire and suggest to use
  flux- or charge qubits for their measurement.  As an application we use the
  notion of fidelity within a first-quantized formalism in order to derive new
  results and insights on the generating function of the full counting
  statistics.
\end{abstract}


\maketitle

\section{Introduction}

Mesoscopic devices exhibit an extraordinary rich and complex behavior;
their proper characterization has sparked numerous ideas.  Two such
basic mesoscopic characteristics are the stability of a quantum system
\cite{peres}, nowadays compiled under the terms `fidelity' or
`Loschmidt echo', and the full counting statistics of charge
transport, expressed through the generating function for the
distribution of charge transmitted across a quantum wire \cite{ll23}.
Here, we combine these items with recent efforts aiming at the
physical realization of quantum bits, the controllable quantum
two-level systems which are the basic elements of a quantum computer
\cite{nielsonchuang}; by now, a number of devices have been realized
in the laboratory, among them particular solid-state implementations
such as flux- \cite{chiorescu} or charge- \cite{vion} qubits which are
easy to couple to. The purpose of this letter is three-fold: i) to
draw attention to the equivalence between the fidelity of a quantum
system and the generating function for the full counting statistics,
two quantities which have been conceived to be unrelated so far. This
insight generalizes the concept of fidelity to mixed states and
many-particle systems. ii) to suggest measuring the fidelity/full
counting statistics using quantum bits by exploiting the induced
`decoherence' as a signal; this is opposite to the standard setting
where the main interest is in the qubit's decoherence due a noisy
environment \cite{makhlin} or a measuring device \cite{averin}.  Our
proposal renders the theoretical concepts of fidelity/full counting
statistics amenable to real experimental tests, e.g., using
high-quality qubits available today \cite{chiorescu,vion}.  In this
context, we recognize earlier suggestions to use two-level
systems/qubits as measuring devices \cite{schoelkopf}. iii) to use the
equivalence between the full counting statistics and the fidelity to
obtain further information on the statistics of charge transport.
\begin{figure}[ht]
  \includegraphics[width=8.0cm]{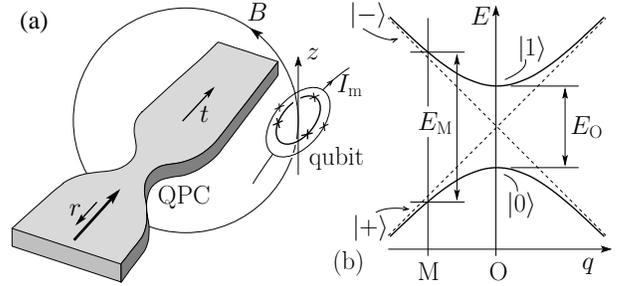}
  \caption[]{(a) Layout: A (flux) qubit (not to scale, placed
    downstream the scatterer) is used to measure the fidelity or full
    counting statistics of a quantum point contact (QPC); the
    measurement current $I_\mathrm{m}$ is switched on only during the
    readout of the qubit state. For a magnetic coupling $\lambda h
    \sigma_z$, the direction of the magnetic field defines the
    $z$-axis.  (b) Level scheme for the qubit, which is initially
    prepared in the ground state $|0\rangle$ at the fully frustrated
    point O and suddenly switched to M at $t=0$ for the measurement,
    producing an initial state $|\Psi\rangle =
    [|+\rangle+|-\rangle]/\sqrt{2}$. The semi-classical states
    $|\pm\rangle$ carry circulating currents perturbing the wire in
    the measurement of the fidelity; in the measurement of the full
    counting statistics the state of the qubit is incrementally
    changed at each passage of a charge through the wire. }
  \label{fig:qubit}
\end{figure}

In this context, there are two points of view in describing a system
(such as a regular or chaotic dot, a quantum wire, etc.) coupled to a
qubit: i) the qubit-centered view, where the dot/wire acts as a noisy
environment producing decoherence of the qubit state \cite{makhlin}
--- this traditional view is pursued in the field of quantum
computing; ii) the dot/wire-centered view, where the qubit serves as a
measurement device providing information on the system (the dot or
wire).  This is the new standpoint we take in the present paper where
we are interested in two system properties, the fidelity and the full
counting statistics.

The notion of {\it fidelity} has been introduced in order to quantify
the stability of a quantum system (described by the Hamiltonian
$H_\mathrm{sys}$) under the action of a small external perturbation
$\lambda h$ \cite{peres}: evaluating the evolution of an initial state
$\Psi$ under the action of the system's Hamiltonian $H_\mathrm{sys}$,
$\Psi(t) = \exp(-iH_\mathrm{sys}t/\hbar) \Psi$, comparison is made
with the perturbed evolution $\Psi_\lambda (t) =
\exp[-i(H_\mathrm{sys}+\lambda h)t/\hbar]\Psi$ through the matrix
element \cite{modsq}
\begin{eqnarray}
   \chi_\mathrm{fid} (\lambda,t) &=& 
   \langle \Psi_\lambda (t)|\Psi(t)\rangle
   \label{chi_fid}\\
   &=&\langle\Psi| e^{i(H_\mathrm{sys}+\lambda h)t/\hbar} 
   e^{-iH_\mathrm{sys}t/\hbar}
   |\Psi\rangle; \nonumber
\end{eqnarray}
a quantum system with a chaotic classical correspondent exhibits a
rapid time-decay of the fidelity $\chi_\mathrm{fid}$
\cite{peres,jalabert}, while a regular classical analogue leads to its
saturation at a finite value \cite{peres}. Above, the fidelity has
been defined for a pure state formulated in a first-quantization
language; the definition (\ref{chi_fid}) agrees with its quantum
information theoretic analogue for two pure state $\Psi$ and
$\Psi_\lambda$ \cite{nielsonchuang}.

In general, the interference between two wave functions as manifested
in (\ref{chi_fid}) is difficult to measure. Here, we put forward the
idea to couple the system under investigation to an external device
(e.g., a two-level system in the form of a spin or a qubit) which
simultaneously acts as a perturbation and as a measuring device for
the fidelity, see Fig.\ \ref{fig:qubit}(a): the coupling $\lambda h
\sigma_z$ entangles the system with the measurement device, thus
transferring information of the system's evolution which then can be
measured along with the quantum state of the device.  A similar idea
has been introduced in the context of the {\it full counting
  statistics} (FCS) which is characterizing the charge transport
through a quantum wire. The task then is to count the number of
transferred charges which corresponds to the measurement of the
integrated current $Q(t)=\int_0^t dt'\, I(t')$. The straightforward
(classical) Ansatz \cite{ll1} for the generating function
$\chi(\lambda,t) = \langle \exp[i\lambda Q(t)] \rangle$ is problematic
as no prescription for the time-ordering of current operators
appearing at different instances of time is given. The idea to couple
the system to a measurement device resolves these problems: following
the work of Levitov and Lesovik \cite{ll23}, the complete statistical
information can be obtained by coupling the wire to a spin degree of
freedom serving as the measuring device; the Fourier coefficients
$\int d \lambda \exp(-i\,n\lambda) \chi_{\rm\scriptscriptstyle
  FCS}(\lambda,t)$ of the generating function
\begin{equation}
   \chi_{\rm\scriptscriptstyle FCS}(\lambda,t) = 
   \mathrm{Tr}_\mathrm{w}
   [\rho^\mathrm{w}(0)\, e^{i(H_\mathrm{w}+\lambda h)t/\hbar} 
   e^{-i(H_\mathrm{w}-\lambda h)t/\hbar}]
   \label{chi_fcs}
\end{equation}
provide the probabilities $P_n(t)$ for the passage of $n$ charges
during the time $t$. Here, $H_\mathrm{w}$ ($\rho^\mathrm{w} (0)$)
denotes the wire's Hamiltonian (density matrix) and $\lambda h$
describes its coupling to the spin; the trace is taken over the wire's
degrees of freedom, while the off-diagonal matrix element has been
evaluated in spin space, see below.

\section{Equivalence between fidelity and FCS} 

The full counting statistics refers to a many-body system cast in a
second-quantized formalism. Still, comparing (\ref{chi_fid}) and
(\ref{chi_fcs}), we immediately note the similarity of the two
expressions: indeed, replacing the arbitrary perturbation in
(\ref{chi_fid}) by the coupling to a spin (or two-level system)
serving at the same time as a perturbation {\it and} as a measurement
device, we arrive at the form in (\ref{chi_fcs}) with the trace over
the wire's degrees of freedom replaced by the quantum average over the
initial state $\Psi$.  In previous discussions \cite{peres,jalabert}
the fidelity of a quantum system has been related to the
chaotic/regular nature of the system; probing such a system with a
measurement device as described above provides the identical
information. The notion of fidelity introduced here generalizes this
concept to quantum systems without classical analogue as well as mixed
state- and many-particle systems; the quantum information theoretic
definition for density matrices \cite{nielsonchuang} is different,
however.

The fidelity/generating function $\chi(\lambda,t)$ is obtained by
coupling a spin (in zero external magnetic field) to the system via a
Hamiltonian of the form $H_\mathrm{int}=\lambda h\sigma_z$ (the
absence of terms $\propto \sigma_{x,y}$ is crucial). The evolution of
the spin then is described by the reduced density matrix
\begin{equation}
   \rho^\mathrm{s}(t) = \mathrm{Tr}
   \bigl[ e^{-i (H_\mathrm{sys} + H_\mathrm{int})t/\hbar} \rho(0) \, e^{i
   (H_\mathrm{sys} + H_\mathrm{int})t/\hbar}\bigr]
   \label{rho_qubit}
\end{equation}
with the initial separable density matrix $\rho(0) =
\rho^\mathrm{sys}(0) \otimes \rho^\mathrm{s}(0)$ and the trace is
taken over the system degrees of freedom without the spin; for the
fidelity (\ref{chi_fid}) we have to replace $\rho^\mathrm{sys}(0) \to
\rho^\Psi(0) = |\Psi\rangle \langle\Psi|$. Evaluating the spin part
first, we subsequently exploit the cyclic property of the trace and
obtain
\begin{eqnarray}
   &&\rho^\mathrm{s}_{\sigma_z',\sigma_z}(t)
   = \rho^\mathrm{s}_{\sigma_z',\sigma_z}(0)
   \label{rho_qubit_ev}\\
   &&\qquad \times \mathrm{Tr} \bigl[\rho^\mathrm{sys}(0) \,
   e^{i(H_\mathrm{sys}+\lambda h\sigma_z)t/\hbar}
   e^{-i(H_\mathrm{sys}+\lambda h\sigma_z')t/\hbar}\bigr].
   \nonumber
\end{eqnarray}
The measurement of the off-diagonal entry $\rho^\mathrm{s}_{-1,1}
(t)$ provides us with the sought-after quantity $\chi(\lambda,t)$.

Below, we will replace the spin degree of freedom by the more
versatile qubit. Thereby, we transfer the measurement of the fidelity
and of the full counting statistics from the realm of a `Gedanken'
experiment to a practical proposal realizable with today's qubit
technology. In this context, we recognize previous steps taken in this
direction in measuring the fidelity of a quantum kicked rotator
\cite{qkr} and in the measurement of higher order correlators
\cite{prober}, see also Ref.\ \cite{lesovik_94} for alternative
theoretical proposals.

\section{Measurement with qubits}

On a technical level, our problem is described by the Hamiltonian (we
formulate the problem for a wire; the extension to other systems is
straightforward)
\begin{equation}
   H = H_\mathrm{w}+H_\mathrm{q}+H_\mathrm{int}
   \label{ham}
\end{equation}
with $H_\mathrm{w}$ the wire's Hamiltonian of which the fidelity
and/or full counting statistics shall be determined,
\begin{equation}
   H_\mathrm{q} = ({\epsilon}/{2})\sigma_z
   -({\Delta}/{2})\sigma_x
   \label{hamqub}
\end{equation}
is the qubit Hamiltonian written in the semi-classical basis
$|\pm\rangle$ (see Fig.\ \ref{fig:qubit}) and
\begin{equation}
   H_\mathrm{int} = \lambda h \sigma_z \label{hamint}
\end{equation}
is the interaction Hamiltonian coupling the qubit to the wire.  For a
magnetic (transverse) coupling, we have the standard form
\begin{equation}
   H_\mathrm{int} = \frac{1}{c} \int_\mathrm{w} dx I_\mathrm{w}
   ({\bf x}) A_x({\bf x}) \sigma_z \label{hamint_t}
\end{equation}
with $I_\mathrm{w}({\bf x})$ the current flowing in the wire and
$A_x({\bf x}) = \int_\mathrm{q} d{\bf l}\cdot\hat{\bf x} \,
|I_\mathrm{q}|/c |{\bf x} - {\bf l}|$ the gauge potential generated by
the qubit current $I_\mathrm{q}$; the coupling constant $\lambda$ is
defined such as to add a phase $\lambda/2 = 2\pi \int_\mathrm{w} dx
A_x(x) /\Phi_0$ to every electron passing the qubit (here, $\Phi_0 =
hc/e$ denotes the unit of flux).  For an electric (longitudinal)
coupling \cite{pilgrambuttiker}, we have the corresponding expression
\begin{equation}
   H_\mathrm{int} = \int_\mathrm{w} dx \rho_\mathrm{w}({\bf x})
   \varphi({\bf x}) \sigma_z \label{hamint_l}
\end{equation}
with $\rho_\mathrm{w}({\bf x})$ the 1D charge density of the wire and
$\varphi({\bf x})= \int_\mathrm{q} d^2 R \, |\rho_\mathrm{q}|/
\varepsilon |{\bf x} - {\bf R}|$ the electric potential generated by
the qubit charge density $\rho_\mathrm{q}({\bf R})$ ($\varepsilon$ is
the dielectric constant); again, we choose a splitting into $\lambda$
and $h$ such that each electron passing the qubit acquires an
additional phase $\lambda/2 = (e/\hbar v) \int_\mathrm{w} dx
\varphi(x)$ ($e>0$; here, $v$ denotes the (typical) electron velocity
and we assume a slowly varying potential, see below for details).

Typical qubits we have in mind are the flux qubit as implemented by
Chiorescu {\it et al.} \cite{chiorescu} or the charge qubit built by
Vion {\it et al.} \cite{vion}. The generic level diagram for these
devices is shown in Fig.\ \ref{fig:qubit}(b): The semi-classical
states $|+\rangle$ and $|-\rangle$ refer to current- (charge-) states
which are the energy eigenstates away from the optimally frustrated
state. Frustration produces mixing with new eigenstates $|0\rangle =
[|+\rangle +|-\rangle]/\sqrt{2}$ and $|1\rangle = [|+\rangle
-|-\rangle]/\sqrt{2}$ at the optimal point O where no currents/charges
appear on the qubit. In order to measure the fidelity/full counting
statistics, the qubit is prepared in the ground state $|0\rangle$ at
optimal frustration (point O in Fig.\ \ref{fig:qubit}(b) with $\Delta
= E_\mathrm{O}$ and $\epsilon = 0$; this corresponds to the spin-state
polarized along the $x$-axis) and then is suddenly switched (at time
$t =0$) to the measuring point M which has to be chosen sufficiently
far away from O to avoid mixing of the semi-classical states (i.e.,
$\Delta = 0$ and $\epsilon = E_\mathrm{M}$, hence the Hamiltonian
(\ref{hamqub}) describes a spin in a magnetic field directed along the
$z$-axis). On the other hand, the point M should be chosen not too
distant away from O in order to avoid the mixing with other levels. At
the end of the signal accumulation (i.e., at time $t$) the state of
the qubit has to be measured in the following manner: rotation of the
qubit state by $-\pi/2$ ($\pi/2$) around the $y$- ($x$-) axis and
subsequent measurement along the $z$-axis provides us with the matrix
elements $\langle \sigma_x\rangle = \mathrm{Tr_q}[\rho^\mathrm{q} (t)
\sigma_x]$ ($\langle \sigma_y\rangle = \mathrm{Tr_q}[\rho^\mathrm{q}
(t) \sigma_y]$) from which the {\it final result} $\chi (\lambda,t) =
\exp (-i E_\mathrm{M} t / \hbar) [\langle \sigma_x\rangle+i\langle
\sigma_y \rangle]$ follows (the phase $\exp (-i E_\mathrm{M} t /
\hbar)$ compensates the trivial time evolution of the qubit in the
finite residual field).

In order to extract the full counting statistics from the generating
function $\chi_{\rm\scriptscriptstyle FCS}(\lambda,t)$, a tunable
coupling between the wire and the qubit is needed.  The tetrahedral
superconducting qubit proposed recently \cite{feigel} lends itself as
a particularly useful measurement device: its doubly degenerate ground
state emulates a spin in zero magnetic field, hence $H_\mathrm{q} =
0$, while a symmetric charge bias $\delta^{\scriptscriptstyle Q}$
produces an interaction Hamiltonian $H_\mathrm{int} \propto
\delta^{\scriptscriptstyle Q} \delta^{\scriptscriptstyle \Phi}
\sigma_z$ which is linear in the magnetic flux
$\delta^{\scriptscriptstyle \Phi}$ (produced by the wire's current)
threading the qubit and easily tunable with $\lambda\propto
\delta^{\scriptscriptstyle Q}$ (note that imposing a flux
$\delta^{\scriptscriptstyle \Phi}$ the tetrahedral qubit also serves
as a {\it tunable} charge detector).  Alternatively, flux-
\cite{chiorescu} or charge- \cite{vion} qubits can be used with a flux
tunable third junction or an electrically tunable capacitance.

Finally, we have to make sure that the individual electrons passing
through the wire are sufficiently coupled to the qubit.  This is
trivially the case for the electric coupling, where a simple
calculation leads to the estimate $\lambda \approx (4 e^2/ \hbar v_F
\varepsilon) \ln(L/d)$ with $v_F$ the Fermi velocity, $L$ the wire's
length, and $d$ its distance from the qubit; with $c/v_F \varepsilon$
of order $10^2$ or larger a $\lambda$-value beyond unity is easily
realized.  The situation is less favorable for the case of magnetic
coupling: associating the qubit with a magnetic dipole $m_\mathrm{q} =
I_\mathrm{q} S/c$ ($S$ denotes the qubit's area), we obtain a coupling
$\lambda \approx 2 \pi \alpha I_\mathrm{q} S/ced$, $\alpha = e^2/\hbar
c$ the finestructure constant. With $I_\mathrm{q}$ of order $1~\mu$A
and $S/d \sim 1~\mu$m, we find a coupling $\lambda \sim 10^{-2}$.
Similar findings apply to the tetrahedral qubit: applying a
homogeneous flux $\delta^\Phi$, the qubit can be used as a tunable
charge detector with large coupling $\lambda \approx (4 E_J C/\hbar
v_F) (\delta^\Phi/\Phi_0) \ln(L/d)$, $C$ denoting the
capacitive coupling between the wire and the qubit.  On the other
hand, applying a symmetric charge bias $\delta^Q$ we find a magnetic
coupling of order $\lambda \approx \alpha (\delta^Q /2e) I_\mathrm{q}
S/2ced$ (we have chosen a typical parameter $E_J/E_C = 10^2$). As may
be expected, the system is easily coupled to the qubit via electric
interaction, while its magnetic coupling is generically weak and has
to be suitably enhanced, e.g., with the help of a flux transformer.

The {\it magnetic} coupling of the flux qubit allows for the
measurement of both the full counting statistics and the system's
fidelity. Going over to the interaction representation (with the
unperturbed Hamiltonian given by the system Hamiltonian), the
generating function assumes the form (cf.\ (\ref{rho_qubit_ev}))
\begin{equation}
   \chi_{\rm\scriptscriptstyle FCS}(\lambda,t) \!=\! \mathrm{Tr} 
   [\rho^\mathrm{sys}(0) \, \widetilde{\cal T} e^{i\frac{\lambda}{\hbar}\!
   \int_0^t \!dt' h(t')} \, 
   {\cal T} e^{i\frac{\lambda}{\hbar}\!
   \int_0^t\! dt' h(t')}],
   \label{chi_int_rep}
\end{equation}
where ${\cal T}$ ($\widetilde{\cal T}$) denote the (reverse) time
ordering operators; as desired, the magnetic coupling then is
proportional to the integrated current (or transferred charge) $Q(t) =
\int_0^t dt' I_\mathrm{w}(t')$; the form (\ref{chi_int_rep}) reveals
the role of $\chi_{\rm\scriptscriptstyle FCS}(\lambda,t)$ as the
generating function for the cumulants of transferred charge.
Alternatively, the qubit can be viewed as a system perturbation and
$\chi_{\rm\scriptscriptstyle FCS}$ assumes the role of a fidelity
$\chi_\mathrm{fid}$.  The {\it electric} coupling to the charge qubit
generates a quantity $\chi_\mathrm{fid}(\lambda,t)$ whose meaning is
predominantly that of a fidelity; on the other hand, for a uniform and
unidirectional charge motion with velocity $v_F$, the time-integrated
charge $\int_0^t dt' \rho_\mathrm{w}(t')$ can be related to the
transferred charge $Q(t)$ via $Q(t)= v_F\int_0^t dt'
\rho_\mathrm{w}(t')$ and thus provides an approximate access to the
full counting statistics at finite voltage for which the electrons
move in a specific direction along the wire.

\section{FCS with wave functions}

Inspired by the equivalence between the fidelity and the generating
function for the counting statistics, we proceed with the calculation
of $\chi(\lambda,t)$ within a first-quantized formalism. In
particular, we choose a system in the geometry of a point contact and
study the fidelity for the case where a wave packet is incident on the
scatterer. This scheme allows for a more elaborate discussion of the
various parametric dependencies of the fidelity/FCS but suffers from
the restriction to a single particle. In order to ameliorate this
limitation, we extend the discussion to a sequence of two incident
wave packets, from which the extrapolation to the many body case can
be performed. 

Consider a wave packet (for $t \to -\infty$ and traveling to the
right)
\begin{equation}
   \Psi_\mathrm{in}(x,t) \equiv \Psi_f(x,t)
   = \int \frac{d k}{2\pi} \, f(k) \, e^{i(kx-\omega_k t)},
   \label{Psi_in} 
\end{equation}
centered around $k_0 >0$ with $\omega_k= \hbar k^2 / 2m$ and
normalization $\int (dk/2\pi) |f(k)|^2 = 1$, incident on a scatterer
characterized by transmission and reflection amplitudes $t_k$ and
$r_k$.  We place the qubit behind the scatterer to have it interact
with the transmitted part of the wave function. The transmitted wave
packet then acquires an additional phase due to the interaction with
the qubit: for a magnetic interaction the extra phase accumulated up
to the position $x$ amounts to $\delta \phi_A(x) = 2 \pi \int^x dx'\,
A_x(x')/\Phi_0$, independent of $k$; as $x \to \infty$ this adds up to
a total phase $\lambda/2= 2\pi \int_{-\infty}^{\infty} dx\,
A_x(x)/\Phi_0$, cf.\ (\ref{hamint_t}). For an electric interaction,
cf.\ (\ref{hamint_l}), the situation is slightly more involved: the
extra phase can be easily determined for a slowly varying
(quasi-classical) potential of small magnitude, i.e., $e|\varphi| \ll
\hbar^2 k_0^2/2m$.  Expanding the quasi-classical phase $\int^x dx' \,
p(x')/ \hbar$ with $p(x) = \sqrt{2m(E+e\varphi(x))}$ to first order in
the potential $\varphi(x)$ yields the phase
$\delta\phi_\varphi(x)=(e/\hbar v)\int^x dx'\, \varphi(x')$ which
asymptotically accumulates to the value $\lambda/2$; its
$v$-dependence is due to the particle's acceleration in the scalar
potential and will be discussed in more detail below. Moreover, note
that $\phi_A$ changes sign for a particle moving in the opposite
direction ($k \to - k$) (i.e., under time reversal) whereas
$\phi_\varphi$ does not. For a qubit placed behind the scatterer both
magnetic and electric couplings produce equivalent phase shifts.
Depending on the state $|\pm\rangle$ of the qubit, the outgoing wave
(for $t \to \infty$)
\begin{eqnarray}
   && \Psi_\mathrm{out}^\pm (x,t)=
   \int \frac{d k}{2\pi} f(k) e^{-i \, \omega_k t}
   \bigl[r_k e^{-i k x} \Theta(-x) \nonumber \\
   && \qquad\qquad\qquad\qquad\qquad 
   + e^{\pm i \lambda/2} t_k e^{i \, k x}\Theta(x)\bigr]
   \nonumber
\end{eqnarray}
acquires a different asymptotic phase on its transmitted part.  The
fidelity is given by the overlap of the two outgoing waves,
\begin{eqnarray}
   && \chi(\lambda,t) = \int dx {\Psi_\mathrm{out}^-}^*(x,t)
       \Psi_\mathrm{out}^+(x,t)    \label{fid_behind}\\
   && \qquad = \int \frac{d k}{2\pi} (R_k + e^{i \lambda} T_k)
   |f(k)|^2 \equiv \langle R\rangle_f + e^{i \lambda} \langle T\rangle_f, 
   \nonumber
\end{eqnarray}
where $R_k= |r_k|^2$ and $T_k= |t_k|^2$ denote the probabilities for
reflection and transmission, respectively, and we have neglected
exponentially small off-diagonal terms involving products $\int dk
f(-k)^* f(k)$.  The result (\ref{fid_behind}) applies to both magnetic
and electric couplings; its interpretation as the generating function
of the charge counting statistics provides us with the two non-zero
Fourier coefficients $P_0 = \langle R\rangle_f$ and $P_1 = \langle
T\rangle_f$ which are simply the probabilities for reflection and
transmission of the particle.  This result agrees with the usual
notion of `counting' those particles which have passed the qubit
behind the scatterer. When, instead, the interest is in the system's
sensitivity, we observe that the fidelity $\chi(\lambda,t)$ lies on
the unit circle only for the `trivial' cases of zero or full
transmission $T=0,~1$, i.e., in the absence of partitioning, or for
$\lambda=2\pi \mathbb{Z}$; the latter condition corresponds to no
counting or the periodic vanishing of decoherence in the qubit. On the
contrary, in the case of maximal partitioning with $\langle R\rangle_f
= \langle T \rangle_f = 1/2$, a simple phase shift by $\lambda = \pi$
makes the fidelity vanish altogether. Hence, partitioning has to be
considered as a (purely quantum) source of sensitivity towards small
changes, as chaoticity generates sensitivity in a quantum system with
a classical analogue.

The result (\ref{fid_behind}) also applies for a qubit placed in front
of the scattering region provided the coupling is of magnetic nature
(for the reflected wave, the additional phases picked up in the
interaction region cancel, while the phase in the transmitted part
remains uncompensated). However, an electric coupling behaves
differently under time reversal and the fidelity acquires the new form
\begin{equation}
   \chi(\lambda,t) =  e^{i \lambda }
   \bigl(e^{i \lambda} \langle R \rangle_f
   + \langle T \rangle_f\bigr).\label{fid_before}
\end{equation}

Next, we comment on the (velocity) dispersion in the electric coupling
$\lambda$: the different components in the wave packet then acquire
different phases. To make this point more explicit consider a Gaussian
wave packet centered around $k_0$ with a small spreading $\delta k \ll
k_0$ and denote with $\lambda_0$ the phase associated with the $k_0$
mode.  The spreading $\delta k$ in $k$ generates a corresponding
spreading in $\delta \lambda \approx \lambda_0 (\delta k/k_0)$ which
leads to a reduced fidelity
\begin{equation}
   \chi(\lambda,t)= \langle R \rangle_f + \langle e^{i \lambda} T 
   \rangle_f \approx R_{k_0} + e^{i \lambda_0 - 
   \frac{\lambda_0^2 (\delta k)^2}{2 k_0^2}} T_{k_0},
   \label{fid_disp}
\end{equation}
where we have assumed a smooth dependence of $T_k$ over $\delta k$ in
the last equation. The reduced fidelity for $T = 1$ is due to the
acceleration and deceleration produced by the two states of the qubit
\cite{ac}. The wave packets passing the qubit then acquire a different
time delay depending on the qubit's state. As a result, the wave
packets become separated in space with an exponentially small residual
overlap for the Gaussian shaped packets.

Next, we consider the case with two wave packets incident on the
scatterer, $\Psi_\mathrm{in} \propto [\Psi_{f_1}(x_1) \Psi_{f_2}
(x_2)\pm (x_1 \leftrightarrow x_2)]\chi_{s/t}(s_1,s_2)$, where
$f_{1,2}$ denote different wave packet amplitudes, cf.\ 
(\ref{Psi_in}), and $\chi_{s/t}$ are the singlet/triplet spin
functions. Placing the qubit behind the scatterer, we obtain the
fidelity
\begin{eqnarray}
   &&\!\!\chi_2(\lambda,t) = \frac{1}{N_\pm} \Bigl[{\prod_{m=1,2}} 
   \int\! \frac{d k_m}{2\pi} |f_m(k_m)|^2 (R_{k_m} + e^{i \lambda} T_{k_m}) 
   \nonumber \\ 
   &&\pm \! {\prod_{m=1,2}} \int \! \frac{d k_m}{2\pi} 
   f^*_m (k_m) f_{n \neq m} (k_m) (R_{k_m} + e^{i \lambda} T_{k_m})\Bigr],
   \label{fid2_behind}
\end{eqnarray}
with the normalization $N_\pm=1\pm |S|^2$, $S=\int (dk/2\pi)$
$f^*_1(k) f_2 (k)$, and we have made use of the fact that all
components of the wave packet propagate in specific directions such
that $\int d k f^*_n (k) f_m (-k) = 0$. The fidelity
(\ref{fid2_behind}) involves two terms, a direct term independent of
the spin/orbital symmetry and an exchange term that can be neglected
if the two wave packets are well separated either in momentum space
(with $k_{0n}$ sufficiently different) or in real space (with $k_0
\delta x \gg 1$). In both of these cases the overlap $S$ vanishes and
we arrive at the final result
\begin{equation}
   \chi_2 (\lambda,t) = [\langle R\rangle_{f_1} 
   + e^{i \lambda} \langle T\rangle_{f_1}]
   [\langle R\rangle_{f_2} + e^{i \lambda} \langle T\rangle_{f_2}],
   \label{fid2_final}
\end{equation}
i.e., the fidelity of the two-particle system is just the product of
the single particle fidelities. The result can be trivially
generalized to the many particle case provided that exchange terms can
be neglected, as is the case for a sequence of ($N$) properly
separated wave packets. The result $\chi_N (\lambda,t) = \Pi_{n=1}^N
[\langle R\rangle_{f_n} + e^{i \lambda} \langle T\rangle_{f_n}]$ then
has to be compared with the previous finding $\chi_V (\lambda,t) = [R +
e^{i \lambda} T]^{e|V|t/h}$ \cite{lc,ll23} calculated at constant
voltage $V$ within a many-body formalism. The validity of the latter
result is restricted to non-dispersive scattering coefficients $T$ and
$R$. Identifying $e|V|t/h$ with the number $N$ of transmitted
particles the two results agree. However, we note that the present
derivation corresponds to a voltage driven many-body setup where
distinct voltage pulses with unit flux $c \int dt \, |V(t)| = \Phi_0$,
each transferring one particle, are applied to the system \cite{ll23}.
The detailed comparison of these two cases, pulsed versus constant
voltage, and the absence of interference terms in the latter case is
an interesting problem for further investigation. The extension of the
fidelity to the many-body case puts additional emphasis on the {\it
  relation between sensitivity and partitioning} (see also Ref.\ 
\cite{agam_00}) as any finite partitioning with $T < 1$ now generates a
fidelity which is vanishing exponentially in time; the quantity
$-\ln[1-4RT\sin^2(\lambda/2)]/\Delta t$ with $\Delta t$ the time
separation between two wave packets/voltage pulses then plays the role
of the Lyapunov exponent in systems with a classical chaotic
correspondent.

In summary, we have unified the themes of fidelity and full counting
statistics and have shown how to exploit qubits for their measurement.
As an application of these ideas, we have recalculated the generating
function for the full counting statistics within a wave packet
approach. This allows for a more in-depth analysis of the counting
problem and helps to shed light on the ongoing discussion \cite{lc}
relating different expressions for high-order correlators to
variations in the experimental setup.

We thank Lev Ioffe for discussions and acknowledge financial support
by the CTS-ETHZ, the MaNEP program of the Swiss National Foundation
and the Russian Science Support Foundation.

\end{document}